\documentclass[pre,aps,
notitlepage,
nofootinbib,
superscriptaddress,floats,floatfix,showpacs,showkeys,
pdflatex10pt]{revtex4}
\usepackage{graphicx}
\usepackage{dcolumn}
\usepackage{bm}
\usepackage{amssymb}
\usepackage{amsmath}
\usepackage{mathrsfs}
\usepackage{ulem}
\usepackage{color}

\keywords{nuclear form; yrast level}
\begin{document}
%\preprint{APS/123-QED}
\title{Two-component self-gravitating isothermal slab models}

\author{Giuseppe Bertin}
\affiliation{{Department of Physics}, {Universit\`{a} degli Studi di Milano}, {Italy},\\giuseppe.bertin@unimi.it,\,\,  orcid 0000-0002-6607-3881}
\author{Francesco Pegoraro}
 \affiliation{Enrico Fermi Department of Physics, University of Pisa, Italy}
\affiliation{INAF-IAPS, Roma, Italy\\
francesco.pegoraro@unipi.it,\,\,  orcid 0000-0002-7216-5491}

\bigskip

\begin{abstract}
We revisit the problem of the isothermal slab (in standard Cartesian coordinates, density distributions and mean gravitational potential are considered to be independent of $x$ and $y$ and to be a function of $z$, symmetric with respect to the $z = 0$ plane) in the context of the general issues related to the role of weak collisionality in inhomogeneous self-gravitating stellar systems. We thus consider the two-component case, that is a system of heavy and light stars with assigned mass ratio ($\mu$) and assigned global relative abundance ($\alpha$; the ratio of the total mass of the heavy and light stars). The system is imagined to start from an initial condition in which the two species are well mixed and have identical spatial and velocity distributions and to evolve into a final configuration in which collisions have generated equipartition and mass segregation. Initial and final distribution functions are assumed to be Maxwellian. Application of mass and energy conservation allows us to derive the properties of the final state from the assumed initial conditions. In general, the derivation of these properties requires a simple numerical integration of the Poisson equation. Curiously, the case in which the heavy stars are exactly twice as massive as the light stars ($\mu = 2$) turns out to admit a relatively simple analytic solution. Although the general framework of this investigation is relatively straightforward, some non-trivial issues related to energy conservation and the possible use of a virial constraint are noted and clarified. The formulation and the results of this paper prepare the way to future studies in which the evolution induced by weak collisionality will be followed either by considering the action of standard collision operators or by means of dedicated numerical simulations.
 \end{abstract}
 
 \pacs{ 98.10.+z, 05.20.Dd, 05.60.Cd}
 
 \keywords{stellar systems,  self-consistency, collisional relaxation, equipartition, mass segregation}

\maketitle

\section{Introduction}\label{sec1}

The study of galaxies as self-gravitating stellar systems often proceeds by means of the Vlasov-Poisson system of equations (e.g., see \cite{cha42}). Smaller stellar systems, such as globular clusters, are known to be affected by weak collisionality (e.g., see \cite{spi87}). In recent years, great progress in telescopes, instruments, and computers is making it possible to test some of the fundamental ideas at the basis of the relevant relaxation processes, either by observations or by dedicated numerical experiments.

Standard relaxation, by means of the cumulative effects of stellar encounters, is expected to be responsible for a number of phenomena that range from isotropization, equipartition, mass-segregation, and evaporation to more curious phenomena such as the gravothermal catastrophe (\cite{lyn68}) or to a rather controversial  instability  associated with mass segregation (\cite{spi69}; \cite{vis78}; \cite{mer81}). In the presence of collisions stars with different masses are expected to be characterized by different orbital evolution. Purely collisionless stellar systems are also subject to various relaxation processes (\cite{lyn67}); however, these processes are expected to lead to equilibria for which the phenomena mentioned above are largely absent. In particular, collisionless stellar systems are generally thought to be well described by a simple one-component distribution function even when they are made of stellar populations characterized by a rather wide spectrum of stellar masses. For such simple stellar systems the connection with the observations is performed by converting the dynamically derived quantities into observable quantities by means of a constant mass-to-light ratio (under the assumption of homogeneity of the stellar population characterizing the system). In contrast, if significant collisional effects occur gradients in the relevant mass-to-light ratios should instead be unavoidable.

Especially in the context of globular clusters, generally modelled as tidally truncated spherical self-gravitating systems with finite mass, great efforts have been made to address the role of weak collisionality in the evolution of stellar systems in the last 60 years (in particular, see \cite{kin65}; \cite{hen71}; \cite{dac76}; \cite{spi87}; and a vast set of references, some of which can be found in the article \cite{tor19}). As often is the case in the astrophysical context, most of the attention has been turned to the astronomical issues and only in fewer articles to the underlying physical mechanisms. Yet, as already indicated by some unexpected findings by \cite{hen71}, self-consistency and collective effects may be the source of surprising results that require clarification. In other words, some caution is generally advised when applying standard expectations, such as isotropization, equipartition, and mass segregation, to self-gravitating inhomogeneous stellar systems. 

In particular, we may note some interesting results that have emerged from recent simulations dedicated to the dynamical evolution of weakly collisional globular clusters. The trend toward equipartition is only partial, even in the central, most relaxed regions (see \cite{tre13}), in the sense that the central velocity dispersion ratio is found to scale as the inverse mass ratio to a power smaller than 1/2 (as would be expected from full equipartition). In addition, most likely as a result of differential collisionality in the system, isotropization is found to be established only in the central parts of the system, which further out slowly builds up (radially-biased) pressure anisotropy profiles that may resemble those resulting from incomplete violent relaxation, known to be relevant to elliptical galaxies (e.g., see \cite{bia17}). Furthermore, there is clear evidence of evolution in the direction of detailed mass segregation (e.g, see \cite{dow10}; Fig.~1 in the article \cite{tor19}). These theoretical results have been found to be largely consistent with rather recent observations, especially some based on the Hubble Space Telescope or the Gaia mission (in this respect, some relevant references are provided by \cite{tor19}).

This short paper is a preparatory piece of work focused on what we think is one of the simplest models that should allow us to follow in detail the evolution of weakly collisional stellar systems. The geometry is that of a slab model, that is an infinite layer considered to be homogeneous on the $(x,y)$ plane, inhomogeneous in the vertical direction $z$, and symmetric with respect to the equatorial plane $z = 0$. The self-gravitating, stellar-dynamical case is known to be associated with simple analytical solutions (\cite{spi42}) that have a direct counterpart in the plasma case (the slab model of a current sheet is often considered as a textbook case for the introduction to magnetic reconnection in plasma physics; see \cite{har62}). For a very recent combined analytical and numerical study of relaxation processes in a one-component 1-dimensional model, see \cite{rou22}.

The problem that we would like to address is the following. Suppose that the system considered is made of two species, light and heavy stars, in different proportions, and imagine as initial condition a situation that is compatible with the solution of a Vlasov-Poisson system of equation, in which the two species are perfectly well mixed and characterized by a common Maxwellian distribution function in terms of specific energy; we are thus assuming here that light and heavy stars have exactly the same velocity distribution (and $z$-independent thermal speed) and the same spatial distribution in the vertical direction. If we now turn on weak collisionality can we predict the slow evolution and separation of the distribution functions characterizing the two species? Would evolution indeed lead to equipartition and mass segregation as anticipated? Would the distribution functions remain quasi-Maxwellian in the course of evolution?
An application of the standard tools (that is, turning on a collision operator and resorting to some analytical model; for example, a Fokker-Planck description) in this simple but inhomogeneous and self-consistent environment is not at all trivial. Most likely, we will have to make use of numerical experiments. In the present preparatory work we simply calculate the properties of the final state by assuming that it follows the standard expectations (the two species will have a different Maxwell-Boltzmann distribution function with the lighter particles characterized by a larger thermal speed and by a more diffuse vertical distribution); in order to determine the properties of such final state from the initial conditions we assume that mass and energy are conserved, that is, that each species keeps the initial integrated surface mass density and that energy is exchanged by collisions between the two species but the total energy is conserved.
As illustrated in the following text, this preparatory work is relatively simple and straightforward, but already raises some interesting issues. It will be the basis of a follow-up paper in which time evolution will be considered explicitly.

\section{Formulation of the Problem}

Consider a self-gravitating slab model, associated with a Maxwellian distribution function

\begin{equation}
f = f (E) = A \exp{(-E/c^2)}~,
\label{df}
\end{equation}

\noindent normalized to the mass density, so that $A = \rho_0/(\sqrt{2 \pi} c)$, and expressed in terms of the specific energy related to the vertical degree of freedom $E = v^2/2 + \Phi(z)$.  For simplicity, we may assume that the three-dimensional velocity distribution is also Maxwellian and characterized by isotropic velocity dispersion. The quantities $A$ and $c$ are positive constants. The coordinate $z$ denotes the inhomogeneity direction of the slab model, which is taken to be symmetric with respect to the equatorial plane $z = 0$. Therefore, $\Phi(z) = \Phi(-z)$; we take $\Phi(0) = 0$, so that $\rho_0$ represents the mass density at $z = 0$. 

The self-gravitating solution is found by solving the Poisson equation

\begin{equation}
\frac{d^2 \Phi}{d z^2} = 4 \pi G \int_{-\infty}^{\infty} f dv~,
\label{poisson}
\end{equation}

\noindent under the regularity boundary condition $d \Phi/ d z = 0$ at $z =0$ and leads to the well-known result

\begin{equation}
\rho = \rho_0 \frac{4}{[\exp{(-z/z_0) + \exp{(+z/z_0)}}]^2} = \rho_0 \exp{( - \Phi/c^2)}~,
\label{cosh}
\end{equation}

\noindent where a measure of the thickness of the slab is given by 

\begin{equation}
z_0^2 = \frac{c^2}{2 \pi G \rho_0}~.
\label{thickness1}
\end{equation}

\noindent If we integrate the density $\rho$ over the vertical coordinate we obtain the total (projected) surface density $\sigma$ and we can easily show that $2 \rho_0 z_0 = \sigma$. Note that, independently of the solution obtained, we may introduce different measures of the thickness. One possibility is that of the length scale $z_m$ defined by the implicit relation

\begin{equation}
\rho(z_m) = \frac{1}{2} \rho_0~.
\label{thickness2}
\end{equation}

\noindent For the self-gravitating solution recorded above we have $z_m \approx 0.88 z_0$. Given the experience with studies of mass segregation in spherical geometry (globular clusters; \cite{dev16}, \cite{tor19}), another possibility is that of the half-mass length $z_{Mi}$ defined as the value of the vertical coordinate at which

\begin{equation}
\int_0^{z_{Mi}} \rho_{i} dz = (1/2) \int_0^{\infty} \rho_i dz~.
\label{thickness3}
\end{equation}

\noindent For the one-component self-gravitating solution recorded above we have $z_M \approx 0.55 z_0$.

Now imagine that the system is made of two species, heavy and light particles, with masses $m_1 > m_2$, and related distribution functions $f_1$ and $f_2$, each of them of the form given in Eq.~(\ref{df}), that is, with coefficients $A_1$, $A_2$, $c_1$, and $c_2$.

In the purely collisionless case, a condition with $A_1 + A_2 = A$, and $c_1 = c_2 = c$ would be a perfectly viable equilibrium solution of the Vlasov-Poisson system of equations if we take for $\Phi(z)$ the result of Eq.~(\ref{cosh}). In particular, the two species would have the same thermal speed ($c_1 = c_2$) and the same spatial distribution (with $z_{m1} = z_{m2} \approx 0.88 z_0$). (This situation is that naturally expected in stellar systems relaxed via collisionless mechanisms; \cite{lyn67}.)

The problem is the following. If we take the condition just stated as an initial condition and we turn on the presence of weak collisionality, is it possible to prove, at least qualitatively, that the time evolution induced by collisions leads to quasi-equilibria with distinct properties, that is $c_1 < c_2$ and $z_{m1} < z_{m2}$ (or the corresponding relation for other measures of the thickness of the two species), consistent with the expectations that the system should tend toward equipartition and mass segregation? It is not obvious that the two distribution functions would evolve by remaining Maxwellian, but we may first start by the assumption that they do, study the evolution of the coefficients $A_i$ and $c_i$, with the related change in time of the self-gravitating solution, and then discuss the merits of this simplifying assumption.

\section{A direct solution}

A first solution to the problem thus posed is obtained by addressing the issue of the {\it final} state that we may presume collisions would tend to achieve asymptotically, in the long run. In other words, we may assume that collisions enforce a Maxwellian with a common temperature (often called a condition of equipartition), so that the final state will be characterized by

\begin{equation}
\frac{c^2_{1f}}{c^2_{2f}} = \frac{m_2}{m_1} \equiv \frac{1}{\mu}~.
\label{equipartition}
\end{equation}

\noindent In addition, during the process we expect the number of particles and the total energy (per unit surface) to be conserved. In other words:

\begin{eqnarray}
A_{1f} (\sqrt{2 \pi} c_{1f}) \int_{-\infty}^{\infty} \exp{[-\Phi_f(z)/c^2_{1f}}]dz  = \rho_{01f} \int_{-\infty}^{\infty} \exp{[-\Phi_f(z)/c^2_{1f}}]dz = \nonumber \\  
= 2 \rho_{01f} z_{01f} = \int_{-\infty}^{\infty} \int_{-\infty}^{\infty} f_{1} dv dz = 2 \rho_{01} z_0 = \sigma_{1}~,
\label{masscons1}
\end{eqnarray}

\begin{eqnarray}
A_{2f} (\sqrt{2 \pi} c_{2f}) \int_{-\infty}^{\infty}  \exp{[-\Phi_f(z)/c^2_{2f}}]dz  = \rho_{02f} \int_{-\infty}^{\infty}  \exp{[-\Phi_f(z)/c^2_{2f}}]dz = \nonumber \\  
= 2 \rho_{02f} z_{02f} =\int_{-\infty}^{\infty} \int_{-\infty}^{\infty}  f_{2} dv dz = 2 \rho_{02} z_0 = \sigma_{2}~,
\label{masscons2}
\end{eqnarray}

\noindent where $\sigma_{i}$ are the surface (projected) densities of the two species. Note that $ z_{01f}$ and $z_{02f}$ are defined by Eqs.~(\ref{masscons1}) - (\ref{masscons2}), whereas the two relations $2 \rho_{01} z_0 = \sigma_{1}$ and $2 \rho_{02} z_0 = \sigma_{2}$ derive directly from the self-consistent solution Eq.~(\ref{cosh}).

We then argue that the transition between the two configurations occurs without change of energy 

\begin{equation}
K_f - K + W_f - W = 0~.
\label{energyconsdiff}
\end{equation}

\noindent Here the relevant kinetic energy $K$ is defined naturally as

\begin{equation}
K = \int_{-\infty}^{\infty}  \int_{-\infty}^{\infty} \frac{1}{2}  v^2 f_1 dv dz +  \int_{-\infty}^{\infty}  \int_{-\infty}^{\infty}  \frac{1}{2}  v^2 f_2 dv dz~.
\label{kineticenergy}
\end{equation}

\noindent Equation (\ref{energyconsdiff}) involves only the difference in gravitational energy between final and initial configuration, which allows us to eliminate possible undesired divergences. In practice, we may refer to the standard definition in terms of the gravitational field so that

\begin{equation}
W_f - W  = - \frac{1}{8 \pi G} \int_{-\infty}^{\infty}  \left[\left(\frac{d \Phi_f}{d z}\right)^2 - \left(\frac{d \Phi}{d z}\right)^2\right] d z~.
\label{gravenergydiff}
\end{equation}

\noindent Note that the initial and the final configuration are required to be characterized by the same (projected) surface density $2 \rho_0 z_0$ and thus by exactly the same (constant) gravitational field at large distances from the equatorial plane; such constant terms cancel out so that no divergence occurs in the last equation.

Equations (\ref{equipartition}), (\ref{masscons1}), (\ref{masscons2}), and (\ref{energyconsdiff}) set four conditions from which we can derive the final values of the four coefficients that define the two distribution functions from the assumed initial values. Note that the self-consistent problem requires a solution for the potential $\Phi_f(z)$ which should be obtained from:

\begin{equation}
\frac{d^2 \Phi_f}{d z^2} = 4 \pi G (\int_{-\infty}^{\infty}  f_{1f} dv + \int_{-\infty}^{\infty}  f_{2f} dv)
\label{poisson2c}
\end{equation}

\noindent with boundary conditions $\Phi_f = 0$ and $d\Phi_f/dz =0$ at $z = 0$.

The one-component problem to determine the self-consistent potential of Eq.~(\ref{cosh}) has two scales (either $A$ and $c$ or $\rho_0 = \rho_{01} + \rho_{02}$ and $z_0$) and no free dimensionless parameters. The two-component problem to determine the final self-consistent potential $\Phi_f(z)$ is a two-parameter problem (the scales can be kept as above); the two dimensionless parameters are the mass density ratio $\alpha \equiv \sigma_{1}/\sigma_{2}$ and the particle mass ratio $\mu \equiv m_1/m_2$ that define the adopted relative properties of the two components.  In principle, this approach can be extended to an arbitrary number $N$ of components in which case there will be two scales and $N-2$ dimensionless parameters.

\subsection{Note on the virial theorem}

The hydrostatic equilibrium condition for the one-component isothermal slab model (in the notation used earlier in this text) is 

\begin{equation}
c^2 \frac{d \rho}{dz} = - \rho \frac{d \Phi}{d z}~.
\label{hydrostatic}
\end{equation}

\noindent The standard procedure to derive the scalar virial theorem is to multiply the equation by z and integrate over the spatial coordinate, that is

\begin{equation}
\int_{-\infty}^{\infty} \left(z c^2 \frac{d \rho}{dz}\right) dz = - \int_{-\infty}^{\infty} \left(z \rho \frac{d \Phi}{d z}\right) dz~.
\label{virial}
\end{equation}

Here we follow the general approach taken when dealing with the energy conservation of Eq.~(\ref{energyconsdiff}) and work on the difference between the two equations that govern the hydrostatic equilibrium in the final and the initial configuration:

\iffalse 
\begin{eqnarray}
\int_{-\infty}^{\infty}  \left(z c_{1f}^2 \frac{d \rho_{1f}}{dz}\right) dz + \int_{-\infty}^{\infty} \left(z c_{2f}^2 \frac{d \rho_{2f}}{dz}\right) dz - \int_{-\infty}^{\infty} \left(z c^2 \frac{d \rho_{1}}{dz}\right) dz +\int_{-\infty}^{\infty}  \left(z c^2 \frac{d \rho_{2}}{dz}\right) dz   \nonumber \\
= - \int_{-\infty}^{\infty}  \left[z (\rho_ {1f} + \rho_{2f})\frac{d \Phi_f}{d z}\right]dz + \int_{-\infty}^{\infty} \left(z \rho\frac{d \Phi}{d z}\right) dz~.~~~~ ~~~~ 
\label{virial2C}
\end{eqnarray}
\fi
\begin{eqnarray}
\int_{-\infty}^{\infty}  \left(z c_{1f}^2 \frac{d \rho_{1f}}{dz} \, + \, z c_{2f}^2 \frac{d \rho_{2f}}{dz}\right) dz - \int_{-\infty}^{\infty} \left(z c^2 \frac{d \rho_{1}}{dz} + z c^2 \frac{d \rho_{2}}{dz}\right) dz  = \nonumber \\
= - \int_{-\infty}^{\infty}  \left[z (\rho_ {1f} + \rho_{2f})\frac{d \Phi_f}{d z}\right]dz + \int_{-\infty}^{\infty} \left(z \rho\frac{d \Phi}{d z}\right) dz~.~~~~ ~~~~ 
\label{virial2C}
\end{eqnarray}
\noindent The left-hand side integrated by parts by treating the individual elements as

\begin{equation}
\int_{-\infty}^{\infty}  \left(z c^2 \frac{d \rho}{dz}\right) dz = - c^2 \int_{-\infty}^{\infty}  \rho dz 
\label{virialK}
\end{equation}

\noindent gives $- 2(K_f - K)$.

The right-hand side can be written as 

\begin{eqnarray}
 - \int_{-\infty}^{\infty}  \left(z \rho_f \frac{d \Phi_f}{d z}\right)dz + \int_{-\infty}^{\infty}  \left(z \rho\frac{d \Phi}{d z}\right) dz =  \nonumber \\
 = - \int_{-\infty}^{\infty} \left[z \frac{1}{4 \pi G} \frac{1}{2}\frac{d}{dz}\left(\frac{d \Phi_f}{d z}\right)^2\right] dz + \int_{-\infty}^{\infty} \left[z \frac{1}{4 \pi G} \frac{1}{2}\frac{d}{dz}\left(\frac{d \Phi}{d z}\right)^2\right] dz = \nonumber \\
+ \frac{1}{8 \pi G}\int_{-\infty}^{\infty} \left[\left(\frac{d \Phi_f}{d z}\right)^2 - \left(\frac{d \Phi}{d z}\right)^2\right] d z = - (W_f - W)~.
\label{virial2Crhs}
\end{eqnarray}

In conclusion, 
energy conservation Eq.~(\ref{energyconsdiff}) can be expressed in terms of only the change in kinetic energy, which must vanish:

\begin{equation}
K_f = K~.
\label{energyconsfin}
\end{equation}

\noindent In other words, once the virial constraint is imposed, the variation of kinetic energy and the variation of gravitational energy must vanish separately.

The kinetic energy (per unit surface) for the initial state is readily calculated 

\begin{equation}
K = \int_{-\infty}^{\infty} \int_{-\infty}^{\infty} \frac{1}{2}  v^2 f_1 dv dz + \int_{-\infty}^{\infty}  \int_{-\infty}^{\infty} \frac{1}{2}  v^2 f_2 dv dz = \rho_0 z_0 c^2 = 2\pi G \rho_0^2 z_0^3~.
\label{kineticenergy2C}
\end{equation}

\section{Initial state and dimensionless formulation}

We may use the initial conditions to set the problem in dimensionless form. In particular, we define the dimensionless potential as 

\begin{equation}
\psi \equiv \frac{\Phi}{c^2}~,
\label{dimlesspot}
\end{equation}

\noindent and the dimensionless vertical coordinate

\begin{equation}
\zeta \equiv \frac{z}{z_0}~,
\label{dimlessz}
\end{equation}

\noindent where $z_0$ is given by Eq.~(\ref{thickness1}). We assume that the two components in the initial state have the same thermal speed $c_1 = c_2 = c$ and are in relative mass proportion with 

\begin{equation}
\alpha \equiv \frac{\rho_{01}}{\rho_{02}} = \frac{\sigma_{1}}{\sigma_{2}} ~,
\label{massprop}
\end{equation}

\noindent where the total mass density in the equatorial plane is given by $\rho_0 = \rho_{01} +  \rho_{02}$. Then the two-component case, with initial conditions $\rho_1(0) = \rho_{01}$, 
$\rho_2(0) = \rho_{02}$, $c_1 = c_2 = c$ is described in dimensional form by the two functions

\begin{equation}
f_1 = \frac{\rho_{01}}{\sqrt{2 \pi} c} \exp{\left[-(v^2/2 + \Phi(z))/c^2\right]}
\label{df1}
\end{equation}

\noindent and 

\begin{equation}
f_2 = \frac{\rho_{02}}{\sqrt{2 \pi} c} \exp{\left[-(v^2/2 + \Phi(z))/c^2\right]}~.
\label{df2}
\end{equation}

\noindent Thus the Poisson equation for the initial state

\begin{equation}
\frac{d^2 \Phi}{d z^2} = 4 \pi G (\int_{-\infty}^{\infty}  f_1 dv + \int_{-\infty}^{\infty} f_2 dv)
\label{poisson2C}
\end{equation}

\noindent in dimensionless form becomes

\begin{equation}
\frac{d^2 \psi}{d \zeta^2} = 2 \exp{(- \psi)}
\label{poisson2Cdimless}
\end{equation}

\noindent with boundary conditions $\psi (0) = 0$, $(d\psi/d\zeta )(0) = 0$.

\section{Final configuration}

For the final configuration, let us introduce the two scale heights $z_{01f}$ and $z_{02f}$ for the two components in such a way that the standard relations with the (projected) surface density $2 \rho_{01f} z_{01f} = \sigma_{1f}$ and $2 \rho_{02f} z_{02f} = \sigma_{2f}$ apply.

If we consider as initial state the one with $c_1 = c_2 = c$ and $A_1 = \alpha A_2$, following the definition of Eq.~(\ref{massprop}), we may refer to the scales $\rho_0$ and $z_0$ and reduce the four conservation constraints to:

\begin{equation}
\frac{c^2_{1f}}{c^2_{2f}} = \frac{1}{\mu}~,
\label{equipartitionD}
\end{equation}

\begin{equation}
2 \rho_{01f} z_{01f} =  \sigma_{1f} = 2 \rho_{01} z_0~,
\label{masscons1D}
\end{equation}

\begin{equation}
2 \rho_{02f} z_{02f} = \sigma_{2f} = 2 \rho_{02} z_0~,
\label{masscons2D}
\end{equation}

\noindent and

\begin{equation}
K = c_{1f}^2 \sigma_{1f} +  c_{2f}^2 \sigma_{2f} = 2 \rho_0 z_0 c^2 ~.
\label{kineticenergy2CD}
\end{equation}

By combining the first three relations into the fourth  and by dividing by $2 \rho_0 z_0$, we get:

\begin{equation}
\frac{\rho_{01}}{\rho_0} c_{1f}^2  +  \frac{\rho_{02}}{\rho_0}c_{2f}^2  = c^2 ~.
\label{transitionrule}
\end{equation}

\noindent In other words the final velocity dispersion of the first component is related to the initial conditions:

\begin{equation}
c_{1f}^2 = \frac{1 + \alpha}{\alpha + \mu} c^2 ~.
\label{transitionruleexpl}
\end{equation}

\noindent Note that if the mass ratio of the single particles $\mu \equiv m_1/m_2$, as defined in Eq.~(\ref{equipartition}), is larger than unity, the above relation requires, as naturally expected, $c_{1f} < c$ for any value of $\alpha$.

The Poisson equation for the final configuration may be written in dimensionless form, by using the scales of the initial state, that is $A$ and $c$, or $\rho_0$ and $z_0$, so that it becomes an equation for the dimensionless potential $\psi_f= \Phi_f/c^2$ in the vertical coordinate $\zeta = z/z_0$, subject to the same boundary conditions as for Eq.~(\ref{poisson2Cdimless})

\begin{equation}
\frac{d^2 \psi_f}{d \zeta^2} = 2 \left[\frac{\rho_{01f}}{\rho_0}\exp{(- a_1 \psi_f)} +  \frac{\rho_{02f}}{\rho_0} \exp{(- a_2\psi_f)}\right]~,
\label{poisson2Cdimless2C}
\end{equation}

\noindent where

\begin{equation}
a_1 = \frac{c^2}{c_{1f}^2} = \frac{\mu + \alpha}{1+ \alpha} 
\label{transition1}
\end{equation}

\noindent and

\begin{equation}
a_2 = \frac{c^2}{c_{2f}^2} = \frac{1}{\mu} \left(\frac{\mu + \alpha}{1+ \alpha}\right)~. 
\label{transition2}
\end{equation}

\noindent The two ``free parameters" $\rho_{01f}/\rho_0$ and $\rho_{02f}/\rho_0$ are eventually determined by imposing the conditions of mass conservation Eqs.~(\ref{masscons1}) - (\ref{masscons2}).

Equation (\ref{poisson2Cdimless2C}) can be integrated once by multiplying both sides by $d \psi_f/dz$:

\begin{equation}
\frac{1}{2}\left(\frac{d\psi_f}{d \zeta}\right)^2 = 2 \left[\frac{\rho_{01f}}{a_1 \rho_0} + \frac{\rho_{02f}}{a_2 \rho_0} - \frac{\rho_{01f}}{a_1 \rho_0}\exp{(- a_1 \psi_f)} - \frac{\rho_{02f}}{a_2 \rho_0} \exp{(- a_2\psi_f)}\right]~,
\label{2Cintegratedonce}
\end{equation}

\noindent where the integration constant has been chosen in order to meet the imposed boundary conditions at $\zeta = 0$. With simpler notation we may rewrite this equation as

\begin{equation}
\frac{1}{2}\left(\frac{d\psi_f}{d \zeta}\right)^2 = 2 \left[b_1 + b_2  - b_1\exp{(- a_1 \psi_f)} - b_2 \exp{(- a_2\psi_f)}\right]~,
\label{2CintegratedonceS}
\end{equation}

\noindent where $a_1 > 0$ and $a_2 > 0$ are assigned parameters whereas $b_1 > 0$ and $b_2 > 0$ are determined by imposing the conditions of mass conservation Eqs.~(\ref{masscons1}) - (\ref{masscons2}):

\begin{equation}
b_1 \int_{0}^{\infty} \exp{(-a_1 \psi_f)} d \zeta = \frac{\sigma_1}{2 a_1 \rho_0 z_0} = \frac{\alpha}{\mu + \alpha} 
\label{massc1}
\end{equation}

\noindent and

\begin{equation}
b_2 \int_{0} ^{\infty}\exp{(-a_2 \psi_f)} d \zeta = \frac{\sigma_2}{2 a_2 \rho_0 z_0} = \frac{\mu}{\mu + \alpha} ~. 
\label{massc2}
\end{equation}

The mass conservation constraints Eqs.~(\ref{masscons1D}) - (\ref{masscons2D}) lead to the following alternative definitions of the parameters $b_1$ and $b_2$ in terms of the two different thicknesses that characterize the two species in the final state:

\begin{equation}
b_1  = \frac{\alpha}{\mu + \alpha} \left(\frac{z_0}{z_{01f}}\right)
\label{massc1thick}
\end{equation}

\noindent and

\begin{equation}
b_2 = \frac{\mu}{\mu + \alpha} \left( \frac{z_0}{z_{02f}}\right)~. 
\label{massc2thick}
\end{equation}

\noindent The last two expressions restate what is recorded in Eqs.~(\ref{massc1}) - (\ref{massc2}) in terms of the definite integrals in the variable $\zeta$. On the other hand, given the choice of initial conditions, with $a_1 = a_2 = 1$, the Gauss theorem and the conservation of mass also require that $b_1 + b_2 = 1$.  Therefore, a simple linear relation is expected to hold in the plane of the inverse dimensionless thicknesses:

\begin{equation}
\frac{\alpha}{\mu + \alpha} \left(\frac{z_0}{z_{01f}}\right) + \frac{\mu}{\mu + \alpha} \left( \frac{z_0}{z_{02f}}\right) = 1~.
\label{masssegreg}
\end{equation}

\noindent The problem is solved by computing the ``eigenvalue" $b_1 = b_1 (\alpha,\mu)$. After this is done, from Eqs.~(\ref{massc1thick}) - (\ref{massc2thick}) we obtain the thickness ratio:

\begin{equation}
\frac{z_{01f}}{z_{02f}} =  \frac{\alpha}{\mu} \frac{1 - b_1}{b_1}~,
\label{masssegreg1}
\end{equation}

\noindent which, for $\mu >1$, is expected to be smaller than unity.

Equation (\ref{2CintegratedonceS}) can be solved by quadrature. However, in the absence of an explicit solution as is available for  the initial configuration, because of the vanishing derivative boundary condition at $\zeta = 0$, it is more convenient to address the integration of the second order ODE Eq.~(\ref{poisson2Cdimless2C}).

In the absence of other strategies to solve this unusual nonlinear integral eigenvalue  problem we may proceed by iteration. From the initial configuration we have an initial seed for the solution $\Phi (z)$ given by Eq.~(\ref{cosh}). For the assigned values of $\alpha$ and $\mu$ we insert this seed solution in Eqs.~(\ref{massc1}) - (\ref{massc2}) and obtain $b_1^{(0)}$ and $b_2^{(0)}$, which we insert in Eq.~(\ref{poisson2Cdimless2C}) to obtain $\Phi^{(1)}(z)$. This allows us to update the parameters $b_1$ and $b_2$. Convergence is expected in a few steps.

\begin{figure} [h!]\center
\includegraphics[width=0.5\textwidth]{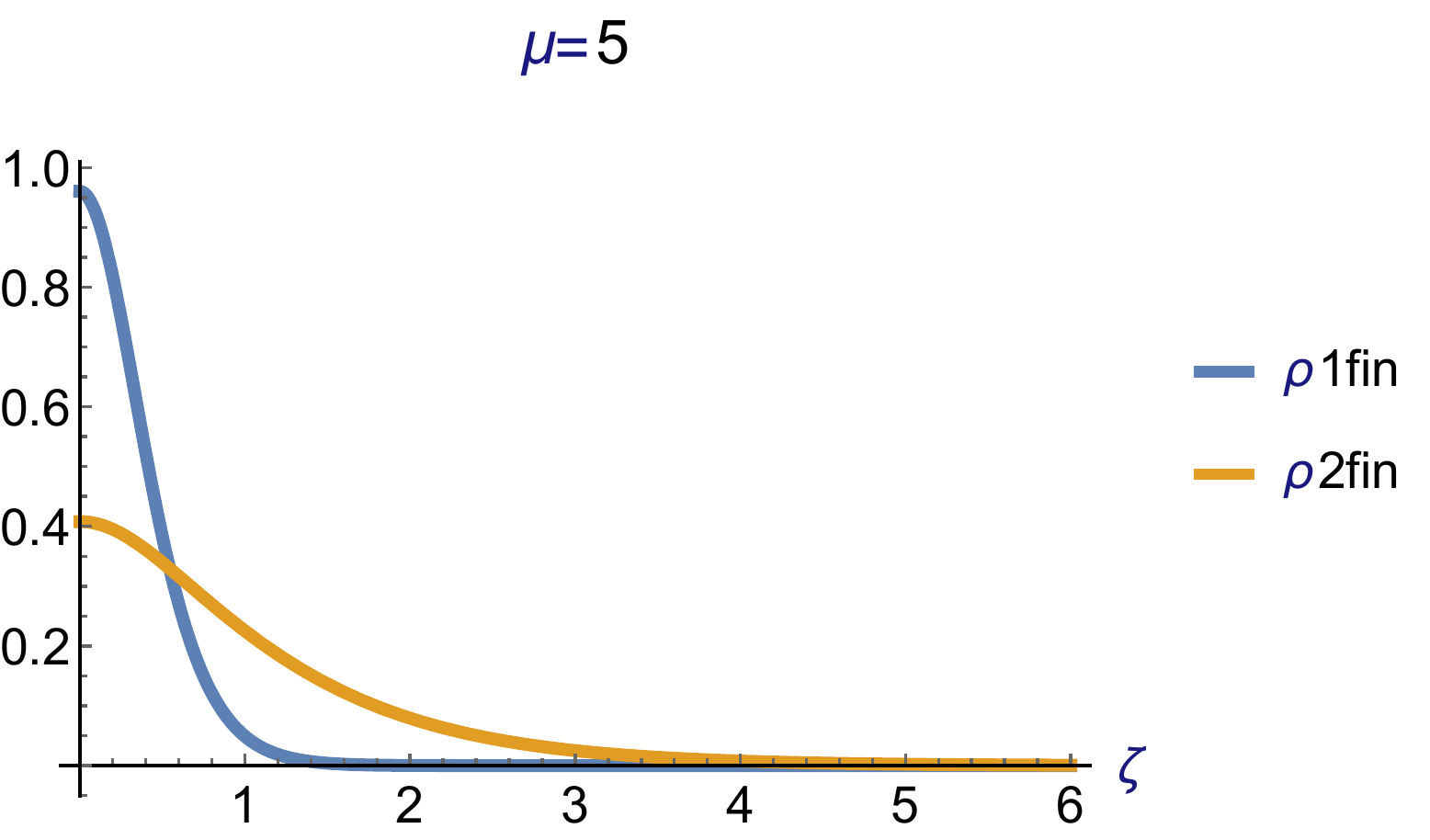}
\caption{Final dimensionless mass density profiles $\rho_{1f}/\rho_0$ and $\rho_{2f}/\rho_0$ for  $\mu = 5$ and $\alpha = 1/2$.}
\label{halfmass1}
\end{figure}

An example of a case with significant mass segregation is shown in Fig.~\ref{halfmass1}, where  the normalized mass density profiles  are plotted versus the normalized variable $\zeta$ for $\mu =5$ and $\alpha = 1/2$.

\begin{figure} [h!]\center
\includegraphics[width=0.5\textwidth]{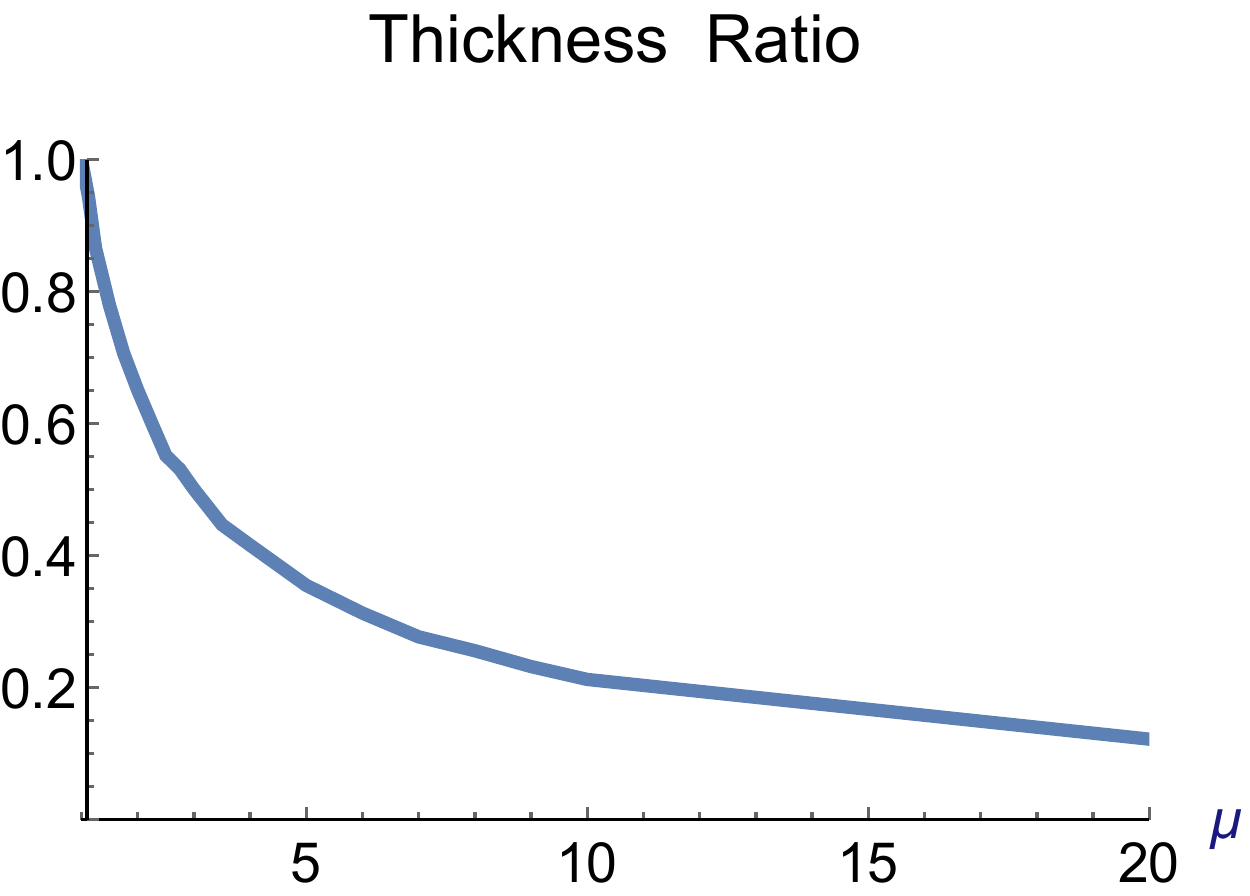}
\caption{Thickness ratio $z_{01f}/z_{02f}$  versus $\mu$ for $\alpha = 1/2$.}
\label{halfmass2}
\end{figure}

The dependence of  the thickness ratio $z_{01f}/z_{02f} $ on the mass  ratio  $\mu$  for fixed $\alpha = 1/2$ is shown  in  Fig.~\ref{halfmass2}. It indicates a rapid decrease of the scale height of the heavier mass component  that tends to flatten out for large values of $\mu$.

\subsection{Subtleties in the definition of the gravitational energy}

In the above discussion we have carried out our argument by referring explicitly to the change in gravitational energy, rather than the direct calculation of the gravitational energy for the initial and for the final configuration. We were prompted to do so by some apparent inconsistencies that would show up if we used other commonly considered definitions of gravitational energy, which can be used safely for problems in which the potential can be set to vanish at infinity (which is not the case here). In particular, we note that, for the initial state defined by Eq.~(\ref{cosh}), the quantity 

\begin{eqnarray}
\frac{1}{2} \int_{-\infty}^{\infty} \rho \Phi dz = \frac{1}{2} \int_{-\infty}^{\infty} [\int_{-\infty}^{\infty} (f_1 +f_2) dv]\Phi dz = \nonumber \\
\rho_0 z_0 c^2 \int_{0}^{\infty}\frac{4}{[\exp{(-\zeta)} + \exp{(+\zeta)}]^2} \psi(\zeta) d\zeta = w \rho_0 z_0 c^2~,
\label{pseudogravenergy}
\end{eqnarray}

\noindent where $w = 2 (1 - \ln{2})\ \approx 0.6137$.  
Taking this as a definition of (positive) gravitational energy would be incompatible with the virial constraint derived earlier in this note.

For systems with finite mass under the condition that the potential vanishes at large distances, an alternative definition of gravitational energy $W_{vir}$ emerges from the standard derivation of the virial theorem, which starts from taking the scalar product of the force term and the position vector (for a general discussion in the context of tensor virial equations, see \cite{cha69}). In our case (see Eq.~(\ref{hydrostatic})), this definition would give  

\begin{equation}
W_{vir} = - \int_{-\infty}^{\infty} z \rho \frac{d \Phi}{dz} dz = - 2 \rho_0 z_0 c^2~.
\label{Wvir}
\end{equation}

\noindent The reason why the two expressions Eq.~(\ref{pseudogravenergy}) and Eq.~(\ref{Wvir}) are different is that in the present 1-dimensional model the potential does not vanish at infinity and some integrations by parts that are normally carried out to relate $W_{vir}$ to other expressions of the gravitational energy cannot be performed here.

Therefore, the procedure adopted in this paper is based on the technique  of computing the difference between the  final  and the initial gravitational energies, as defined in terms of the integral of the energy density given by the square  of the gravitational fields, in such a way as to cancel the undesired divergent terms.

This can also be checked on the final solution that we provide. Starting from Eq.~(\ref{poisson2Cdimless}) 
we obtain 
\begin{equation} 
\frac{1}{2}\left(\frac{d\psi}{d \zeta}\right)^2 = 2 \left[1 -  \exp{(-\psi})\right]~.
\label{2CintegratedonceS energy}
\end{equation}
Then,  from Eq.~(\ref{2CintegratedonceS}), we have
\begin{eqnarray}
\frac{1}{2}\int _0 ^\infty d \zeta \left[ \left(\frac{d\psi_f}{d \zeta}\right)^2 - \left(\frac{d\psi}{d \zeta}\right)^2\right]
=  ~~~~~~~\nonumber 
\\ -  2\int _0 ^\infty d \zeta \left[  b_1\exp{(- a_1 \psi_f)}  + b_2 \exp{(- a_2\psi_f)} -   \exp{(-\psi})\right]. \label{2CintegratedonceS deltaenergy}
\end{eqnarray}
Finally,  after   inserting  Eqs.~(\ref{massc1}) -  (\ref{massc2}) into the r.h.s. of Eq.~(\ref{2CintegratedonceS deltaenergy}), we find 
\begin{equation} 
\frac{1}{2}\int _0 ^\infty d \zeta \left[ \left(\frac{d\psi_f}{d \zeta}\right)^2 - \left(\frac{d\psi}{d \zeta}\right)^2\right]
=0~. 
\label{2CintegratedonceS deltaenergy2}
\end{equation}

\subsection{The special case $\mu = 2$}
   
\begin{figure} [h!]\center
\includegraphics[width=0.5\textwidth]{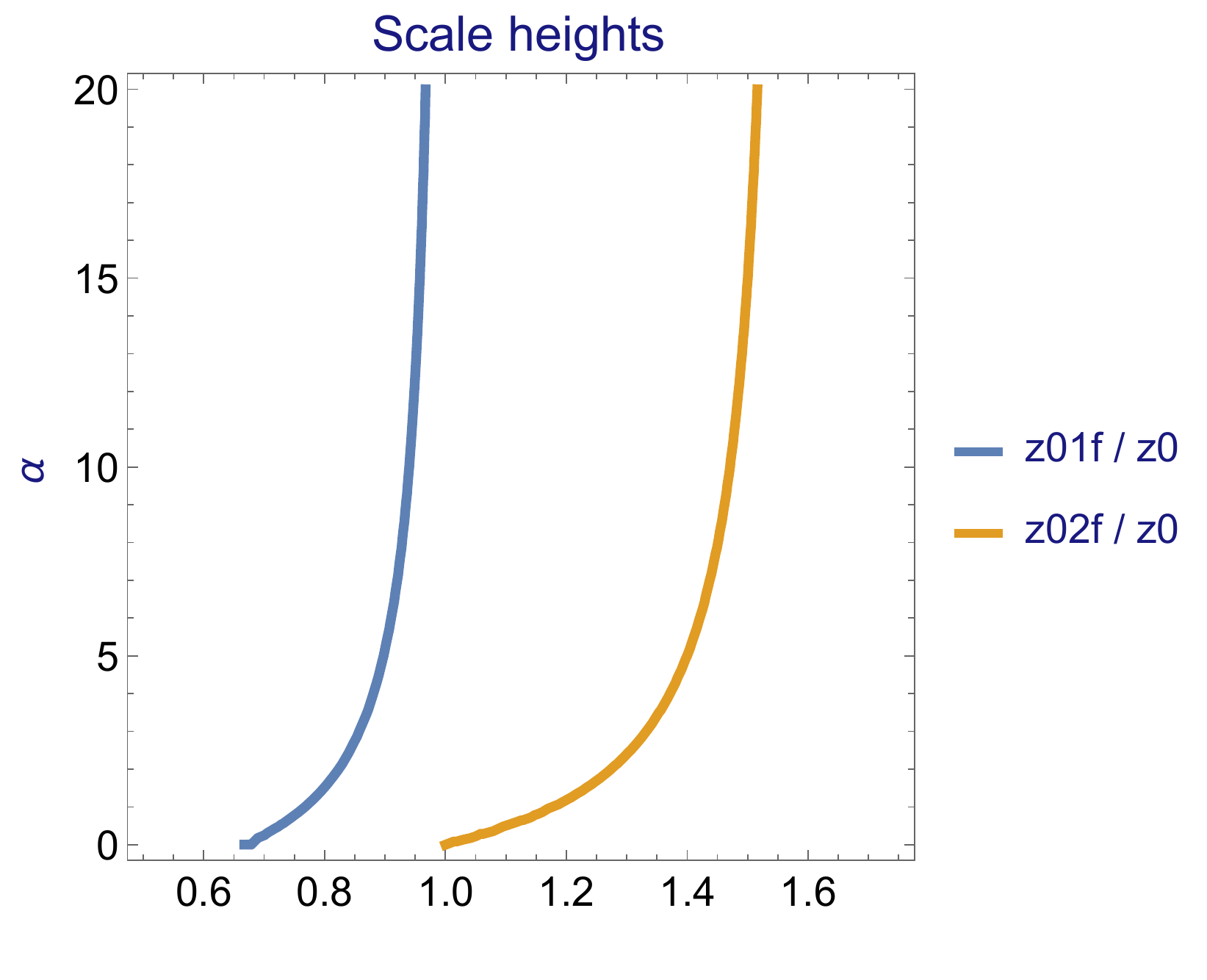}
\caption{Dependence of  the normalized thickness scales  ${z_{01f}/z_0}$  and ${z_{02f} /z_0}$  on $\alpha$ for $\mu = 2$.}
\label{halfmass3}
\end{figure}

   For $\mu = 2$  and free $\alpha$  Eq.~(\ref{massc2})  can be solved for $b_2$ (with $b_2 = 1 -b_1$)  by performing  an elementary integration in terms of the variable $\exp{(-a_2 \psi_f)} $ and  using Eq.~(\ref{2CintegratedonceS})  for the change of variables.  We obtain     \begin{equation} \label{mu2}  b_2  \int_{0}^{\infty} \exp{(-a_2 \psi_f)} d \zeta =  
 \frac{b_2}{2 a_2}\int_{0}^{1} \frac{d\xi }{\left[1  - b_2\xi -(1-b_2) \xi^2\right]^{1/2}} = \frac{1}{1+\alpha/2}~,\end{equation}
   which  does not require the explicit knowledge of $\psi_f$. Recalling that $a_2 =  (1+ \alpha/2)/(1+\alpha)$ and performing the integration over $\xi$, we find the following implicit relationship between $b_2$ and $\alpha$
   \begin{equation} 
 b_2  \,  \frac{ \pi/2  - \arcsin{[b_2/(2-b_2)]} }{(1-b_2)^{1/2}}= \frac{2}{1+\alpha}~. \end{equation}
Using this relationship in Eqs.~(\ref{massc1thick}, \ref{massc2thick}) we can draw,  see Fig.~\ref{halfmass3},  the dependence of  the normalized thickness scales ${z_{01f}/z_0}$  and ${z_{02f} /z_0}$ on  the mass density ratio  $\alpha$ for $\mu= 2$.

For small $\alpha$ most of the mass is in the light component  and, as expected, ${z_{02f} /z_0} \sim 1$,  whereas the  smaller, heavier component  is segregated at ${z_{01f} /z_0} < 1$. For large values of  $\alpha$ mass segregation is reversed with  ${z_{01f} /z_0} \sim 1$ and ${z_{02f} /z_0}  >1$.

\section{Concluding remarks}\label{sec13}

In this preparatory paper, in a simple 1-dimensional model, we have formulated the problem of calculating the final self-gravitating configuration subject to equipartition and mass segregation as expected to be induced by collisional processes starting from an initial condition of well-mixed stellar populations with identical spatial distributions and velocity dispersions. The problem has been solved for the simple case of two components, light and heavy stars. The case in which the heavy stars are twice as massive as the light stars has been shown to admit a simple analytic solution. The adopted procedure can be generalized to the case of N components with N different masses.

This short paper opens the way to a comparison with either analytical investigations or numerical simulations in which the evolution induced by collisionality is followed explicitly, in the context of self-consistent stellar dynamics (the Vlasov-Poisson system of equations, modified by the presence of weak collisions). A similar analysis might then be carried out in the different environment of spherical stellar systems, to bring out analogies and differences that are likely to occur when a different geometry is considered. Indeed, the interest in the present straightforward study is partly due to the fact that some subtleties in the definition of gravitational energy and in the derivation of the virial constraint for the adopted 1-dimensional problem have been identified and discussed. For the spherical case, some curious properties in the solution of a problem similar to the one addressed here had been noted earlier (N.C. Amorisco and L. Ciotti, 2010, private communication).

\begin{acknowledgments}
We thank Luca Ciotti for a number of interesting conversations on the subject of this paper. GB wishes to thank the Department of Physics of the Universit\`{a} di Pisa for the hospitality extended to him when part of this work was carried out.
\end{acknowledgments}


\begin{thebibliography}{99}

\bibitem{bia17}P. Bianchini, A. Sills, M. Miholics, Mon. Not. Roy. Astron. Soc. 471, 1181 (2017)

\bibitem{cha42}S. Chandrasekhar, Principles of Stellar Dynamics (University of Chicago Press, Chicago, 1942) 

\bibitem{cha69}S. Chandrasekhar, Ellipsoidal Figures of Equilibrium (Yale University Press, New Haven, 1969)

\bibitem{cli21}D. Cline, Variational Principles in Classical Mechanics (3rd ed., River Campus Libraries, Rochester, 2021)

\bibitem{dac76}G.S. Da Costa, K.C. Freeman, Astrophys. J. 206, 128 (1976)

\bibitem{dev16} R. de Vita, G. Bertin, A. Zocchi,  Astron. Astrophys.  590, id.A16 (2016)

\bibitem{dow10}J.M.B. Downing, M.J. Benacquista, M. Giersz, R. Spurzem, Mon. Not. Roy. Astron. Soc. 407, 1946 (2010)

\bibitem{har62}E.G. Harris, Il Nuovo Cimento 23, 115 (1962)

\bibitem{hen71}M. H\'{e}non, Astrophys. Space Sci. 13, 284 (1971)

\bibitem{kin65}I. R. King, Astron. J. 70, 376 (1965)

\bibitem{lyn67}D. Lynden-Bell, Mon. Not. Roy. Astron. Soc. 136, 101 (1967)

\bibitem{lyn68}D. Lynden-Bell, R. Wood, Mon. Not. Roy. Astron. Soc. 138, 495 (1968)

\bibitem{mer81}D. Merritt, Astron. J. 86, 318 (1981)

\bibitem{rou22}M. Roule, J.-B. Fouvry, C. Pichon,  P.-H. Chavanis, arXiv:2204.02834 (2022)

\bibitem{spi42}L. Spitzer Jr., Astrophys. J. 95, 329 (1942)
 
\bibitem{spi69}L. Spitzer Jr., Astrophys. J. Letters 158, L139 (1969)

\bibitem{spi87}L. Spitzer  Jr., Dynamical Evolution of Globular Clusters (Princeton University Press, Princeton, 1987)

\bibitem{tor19}S. Torniamenti, G. Bertin, P. Bianchini, Astron. Astrophys. 632, id.A67 (2019)

\bibitem{tre13}M. Trenti, R. van der Marel,  Mon. Not. Roy. Astron. Soc. 435, 3272 (2013)

\bibitem{vis78}E.T. Vishniac, Astrophys. J. 223, 986 (1978) 

\end{thebibliography}
 \end{document}